\documentclass[journal,final,twoside]{IEEEtran}
\usepackage{soul}
\usepackage{graphicx}

\newcommand{\iint}{\int\int}

\begin{document}

\title{Coupled Mode Theory of Optomechanical Crystals}

\author{Sina~Khorasani
\thanks{The author gratefully acknowledges financial support of this work by Laboratoire de Photonique et de Mesure Quantique (LPQM) at EPFL. This work has been supported in part by Iranian National Science Foundation under grant 93026841.}
\thanks{S. Khorasani is currently visiting \'{E}cole Polytechnique F\'{e}d\'{e}rale de Lausanne (EPFL), on leave from School of Electrical Engineering, Sharif University of Technology (e-mail:sina.khorasani@epfl.ch; khorasani@sina.sharif.edu).
}
}

\maketitle

\begin{abstract}
Acousto-optic interaction in optomechanical crystals allows unidirectional control of elastic waves over optical waves. However, as a result of this nonlinear interaction, infinitely many optical modes are born. This article presents an exact formulaion of coupled mode theory for interaction between elastic and photonic Bloch waves moving along an optomechanical waveguide. In general, an optical wavefront is strongly diffracted by an elastic wave in frequency and wavevector, and thus infinite modes with different frequencies and wavevectors appear. We discuss resonance and mode conversion conditions, and present a rigorous method to derive coupling rates and mode profiles. We also find a conservation law which rules over total optical power from interacting individual modes.  We present application examples to the theory to optomechanical waveguides and cavities, as well as non-reciprocal transmission of light and optomechanical switches.
\end{abstract}


\section{Introduction}

\IEEEPARstart{A}{n}  \textit{optomechanical} crystal is essentially a periodic photonic crystal \cite{1} in which electromagnetic waves may propagate at certain frequencies and directions, while mechanical elastic waves are also launched simultaneously. The mechanical waves obey the elastic Bloch properties of the same medium being treated as a periodic phononic crystal. The photo-elastic interaction \cite{2,3,4} between an elastic wave with frequency $\Omega$ and electromagnetic wave at a much higher frequency $\omega$ leads to a type of nonlinear phenomenon which produces new harmonics as $\omega\pm n \Omega$. While both types of waves propagate in the same type of periodic medium, and observe similar Bloch-periodicity and orthogonality conditions \cite{5}, their existence are connected through the photo-elastic tensor \cite{2,4} which contributes through the constitutive relations for electric field and displacement vectors.

In piezo-electric media, low frequency electromagnetic waves may strongly couple and co-exist with their elastic waves, giving rise to a different type of piezo waves, and hence $\omega=\Omega$. Such linear waves may also observe Bloch-periodicity in periodic background, and generate a bandstructure \cite{6} exhibiting wide bandgaps. This is different from the scope of this paper in which optical and elastic waves interact nonlinearly.

This field was brought to the attention by a landmark paper in 2003 \cite{7}, which reported strong Doppler shifting of reflecting electromagnetic waves from an acoustic shock wave. This rapidly opened up numerous applications in the area of phononic crystals \cite{8}, while the recent applications of optomechanics has gone well beyond of classical regime, and quantum mechanical phenomena are being studied. These include manipulation of photo-luminescence \cite{9}, spin control in diamond Nitrogen vacancies for quantum computing purposes \cite{10}, tuning photonic crystals with phonons \cite{11}, and coherent wavelength conversion \cite{12}. Current literature in this area is now quite vast, which is reviewed in several extensive works \cite{13,14}.

The main difficulty with the numerical study of this area arises from the many orders of magnitude difference between the optical $\omega$  and acoustic $\Omega$ frequencies, which is typically being on order of $10^{5}$ to $10^{7}$. Hence, time-domain methods such as nonlinear Finite-Difference Time-Domain \cite{14a,14b} developed for second-order and third-order nonlinear interactions are essentially inefficient to reproduce useful results. The reason is that the total computational time should be at least of the order of $10^{3}$ to $10^{4}$ acoustic time-cycles, which considering the very tiny time-steps being required for ultrashort optical cycles, demands a very huge number of total time steps. For a typical simulation this number of total time-steps could easily blow up to anywhere from $10^{9}$ to $10^{12}$, or even more. This is well beyond the capability of a general purpose personal computer, unless the elastic wave is supposed to be frozen at relatively large time intervals. This method has been implemented by some authors \cite{14c,14d}, but is unable to take care of the frequency differences of Doppler shifts caused by the acousto-optic interaction.

It had been assumed that mechanical loss is a major obstacle to propagate elastic waves inside phononic waveguides over long distances, however, with the recent progress in understanding of propagation of mixed bulk-surface elastic waves \cite{15}, hope for near-lossless propagation in slabs has come into the light. Interestingly, for all practical purposes, two-dimensional slabs of periodic media are being widely studied \cite{16} since true three-dimensional structures cannot exist. Examples include optomechanical interactions in thin membranes such as single-layer graphene \cite{17,18,19,20} and ultrathin high-stress $\textrm{Si}_{3}\textrm{N}_{4}$ membranes \cite{21,21a}. Recent progress in fabrication of ultrathin $\textrm{Si}_{3}\textrm{N}_{4}$ membranes \cite{21,21a} has resulted in record long-lived confined modes, reaching room-temperature quality factors in excess of $10^8$ at $\Omega=1\textrm{MHz}$ \cite{21b}. This remarkable achievement has demonstrated the possibilities for strong photon-phonon interactions over large time intervals, even at elevated temperatures. Enhanced scattering in the subwavelength regime in Si nanostructures is being also shown and pursued \cite{21c}.

It is the purpose of this paper to develop a full and exact theoretical understanding of the nonlinear interaction between elastic and electromagnetic waves from a classical point of view. Without loss of generality, we assume simple dielectric, that is being nonmagnetic, linear, isotropic, and lossless. Using a harmonic-balance method we identify and equate various harmonics propagating at different frequencies and directions. We obtain a set of coupled-mode equations, comparable to those known in standard coupled-mode theory \cite{21d,21e,21f}, which may be solved accurately at discretion, obtaining coupling rates or coupling lengths. Our results confirm the earlier findings \cite{22} based on the boundary perturbation theory \cite{23}. We also present an application example to non-reciprocal transmission of light inside an optomechanical waveguide.

\section{Theory}
The photo-elastic interaction is described using the equation \cite{2,3,4}

\begin{eqnarray}
\label{eq1}
    \Delta \eta_{ij}=p_{ijmn}S_{mn},\\
\nonumber
T_{mn,n}=(c_{mnij}S_{ij})_{,n}=-\Omega^2\rho u_{m},
\end{eqnarray}

\noindent
where $p_{ijmn}$ are the elements of the fourth-rank photo-elastic tensor, $\eta_{ij}$ are elements of the impermeability tensor, and $S_{mn}=\frac{1}{2}(u_{m,n}+u_{n,m})$ are strain components obtained from the displacement vector field $\textbf{u}$. Furthermore, $\rho$, $T_{mn}$ and $c_{mnij}$ are respectively the mass density, and the elements of the stress and stiffness tensor. Symmetry considerations \cite{4} require that $p_{ijmn}=p_{jimn}=p_{ijnm}$ to maintain $S_{mn}=S_{nm}$ and $\eta_{ij}=\eta_{ji}$.

We here adopt a dimensionless system of variables for Maxwell's equations, where $\epsilon_0$ and $\mu_0$ are dropped and the light velocity $c=1$ is normalized. Furthermore, the lattice constant $a$ of the periodic medium is set to unity and all frequency and length measures are correspondingly normalized. Since the medium is taken to be simple, then unperturbed permittivity tensor is simply $[\epsilon]=\epsilon(\textbf{r}) [\textrm{I}]$, where $\epsilon(\textbf{r})=n^2(\textbf{r})$ is the position-dependent squared refractive index. Evidently, this is going to be a biperiodic function of position as $\epsilon(\textbf{r})=\epsilon(\textbf{r}+\textbf{R})$ with $\textbf{R}$ being any discrete lattice vector.

The change in the permittivity of bulk due to a strain field is thus

\begin{equation}
\label{eq2}
[\Delta \epsilon]=(\epsilon^{-1}[\textrm{I}]+[\Delta\eta])^{-1}-\epsilon[\textrm{I}].
\end{equation}

\noindent
If the perturbation is sufficiently small, then (\ref{eq2}) may be approximated as

\begin{equation}
\label{eq3}
[\Delta \epsilon]\approx -[\epsilon][\Delta \eta][\epsilon]=-\epsilon^2[\Delta \eta],
\end{equation}

\noindent
or

\begin{equation}
\label{eq4}
\Delta \epsilon_{ij}\approx -\epsilon_{il}p_{lkmn}S_{mn}\epsilon_{kj}=-\epsilon^2 p_{ijmn}S_{mn}.
\end{equation}

The strain field $[S]=[S(\textbf{r},t)]$ is actually a function of position and time, corresponding to a moving elastic wave at frequency $\Omega$, and therefore $[\Delta \epsilon]=[\Delta \epsilon(\textbf{r},t)]$. It is here appropriate to decompose $[\Delta \epsilon]$ as

\begin{equation}
\label{eq5}
[\Delta \epsilon]=\frac{1}{2}[\delta \epsilon](e^{j\Omega t}+e^{-j\Omega t}),
\end{equation}

\noindent
in which $[\delta \epsilon]$ is a time-independent, but position-dependent symmetric real tensor.

We may now proceed with the Maxwell's equations

\begin{eqnarray}
\label{eq6}
\nabla \times \vec{ \textbf{E}}(\textbf{r},t)=-\frac{\partial}{\partial t} \vec{\textbf{ H}}(\textbf{r},t),\\
\label{eq7}
\nabla \times \vec{ \textbf{H}}(\textbf{r},t)=+\frac{\partial}{\partial t} \vec{\textbf{D}}(\textbf{r},t),
\end{eqnarray}

\noindent
where $\vec{\textbf{D}}=[\epsilon]\vec{\textbf{E}}$. Plugging (\ref{eq5}) in (\ref{eq6}), (\ref{eq7}) and some algebraic simplification gives the coupled set of balanced harmonic equations for the complex phasor vector fields $(\textbf{E}_n,\textbf{H}_n)$, being here referred to as modes, at the frequency component $\omega_n=\omega+n\Omega$ as

\begin{eqnarray}
\label{eq8}
\nabla \times \textbf{E}_n (\textbf{r})=-i\omega_n\textbf{H}_n(\textbf{r})\\
\label{eq9}
\nabla \times \textbf{H}_n (\textbf{r})=+i\omega_n\textbf{D}_n(\textbf{r})
\end{eqnarray}

\noindent
in which the displacement field of the harmonic $n$ is given by

\begin{equation}
\label{eq10}
\textbf{D}_n=\epsilon \textbf{E}_n+\frac{1}{2}[\delta \epsilon](\textbf{E}_{n+1}+\textbf{E}_{n-1}),
\end{equation}

\noindent
with the dependence on position vector $\textbf{r}$ dropped for convenience.

It is here furthermore understood that the modes are not time- or spatial-harmonics of the same frequency as opposed to the theory of nonlinear optics \cite{2}, diffraction theory of gratings \cite{21d}, or multi-port networks \cite{21e,21f}, since in general $\Omega$ is not an integer divisor of $\omega$. The total fields may be now recovered from the summation over individual modes as

\begin{eqnarray}
\label{eq10a}
\vec{ \textbf{E}}(\textbf{r},t)=\sum_{n} \Re[ \textbf{E}_n (\textbf{r})e^{i\omega_n t}],\\
\vec{ \textbf{H}}(\textbf{r},t)=\sum_{n} \Re[ \textbf{H}_n (\textbf{r})e^{i\omega_n t}].
\end{eqnarray}

\noindent
These constitute the total incident and scattered optical fields due to the nonlinear mixing of optical and elastic waves at the original frequencies $\omega$ and $\Omega$, respectively.

\subsection{Coupling coefficients}
It is possible to make use of the vector identity $\textbf{A}\cdot\nabla\times\textbf{B}=\textbf{B}\cdot\nabla\times\textbf{A}-\nabla\cdot(\textbf{A}\times\textbf{B})$, multiplying (\ref{eq8}) and (\ref{eq9}), respectively with $\textbf{H}_n^*$ and $\textbf{E}_n^*$, and subtracting to obtain the master power equation

\begin{equation}
\label{eq11}
\nabla\cdot \bar {\textbf S}_n=\omega_n \Im[U_{en}+U_{mn}].
\end{equation}

\noindent
Here, $\bar {\textbf S}_n=\frac{1}{2}\Re[\textbf{E}_n\times\textbf{H}_n^*]$ is the time-averaged Poynting's vector, and $U_{en}=\frac{1}{2}\textbf{E}_n^*\cdot\textbf{D}_n$ and $U_{mn}=\frac{1}{2}\textbf{H}_n^*\cdot\textbf{H}_n$ are respectively the time-averaged electric and magnetic energy densities of the $n$th harmonic. Integration of (\ref{eq11}) over the crystal's unit cell will give rise to the coupling coefficients due to the  photoelastic, or the so-called bulk electrostrictive forces \cite{23b} as demonstrated below.

One may view the sharp boundaries across dielectrics be softened by a gradual variation of mass density and stiffness within a given virtual transition thickness $t$, while assuming $t$ ultimately approaches zero. Similarly, ultra-high vacuum or low-pressure gas may be modeled by a material with vanishing mass density. This will help to avoid inclusion of extra terms caused by divergences across the boundaries. Otherwise, these divergences need to be treated as separate surface integrals, known as the contribution from radiation pressure \cite{14c}. In general for the case of nano-structured materials with piezoelectric property such as GaAs, radiation pressure terms are negligible compared to the bulk electrostrictive forces \cite{23b}. For non-peizoelectric media, radiation pressure and electrostriction may contribute comparably, or at least within the same order of magnitude \cite{14c}.

We here suppose that the optomechanical crystal is having a properly designed line defect, which allows guided propagation of elastic and optical waves along the direction $\pm x$. A unit cell of such waveguide covers the domain $[0,1]\times(-\infty,\infty)$ and is here denoted by $\textrm{UC}$. Now, (\ref{eq11}) may be integrated over a unit cell to obtain

\begin{equation}
\label{eq12}
I_n(1)-I_n(0)=\omega_n \iint_{\textrm{UC}} \Im [U_{en}+U_{mn}] d^2r,
\end{equation}

\noindent
where $I_n(x)$ denotes the total optical power of the $n$th harmonic propagating along the direction $+x$. Expansion of the integrand here using (\ref{eq10}) gives

\begin{equation}
\label{eq13}
\Im [U_{en}+U_{mn}]=\frac{1}{2}\Im[\textbf{E}_n^*\cdot[\delta \epsilon](\textbf{E}_{n+1}+\textbf{E}_{n-1})],
\end{equation}

\noindent
since $\Im[\textbf{H}_n^*\cdot\textbf{H}_n]=0$, and also $\Im[\textbf{E}_n^*\cdot[\epsilon]\textbf{E}_n]=0$ by transpose symmetry of $[\epsilon]$. This will simplify (\ref{eq12}) as

\begin{equation}
\label{eq14}
I_n(1)-I_n(0)= \frac{\omega_n}{2} \Im \iint_{\textrm{UC}} \textbf{E}_n^*\cdot[\delta \epsilon](\textbf{E}_{n+1}+\textbf{E}_{n-1}) d^2r.
\end{equation}

\noindent
We may now define the coupling rate $\alpha_n$ and coupling length $L_n=\alpha_n^{-1}$ as $I_n(1)=I_n(0)e^{-\alpha_n}$. Hence, if we have normalized the $n$th harmonic power as $I_n(0)=1$, then for the case of $L_n>>1$ we have

\begin{equation}
\label{eq15}
\alpha_n\approx -\frac{\omega_n}{2} \Im \iint_{\textrm{UC}} \textbf{E}_n^*\cdot[\delta \epsilon](\textbf{E}_{n+1}+\textbf{E}_{n-1}) d^2r,
\end{equation}

\noindent
subject to the normalization

\begin{equation}
\label{eq16}
I_n(0)= \int_{-\infty}^{+\infty} \bar {\textbf S}_n\cdot{\hat x} dy=1,
\end{equation}

\noindent
for all $n$. The result (\ref{eq15}) may be compared to the  expression which can be constructed in a similar way by only taking the effect of radiation pressure into account \cite{22} and ignoring the electrostriction as

\begin{equation}
\label{eq16a}
\alpha_n\approx -\frac{\omega_n}{2} \Im \oint_{\textrm{UC}} (\textbf{E}_{n\parallel}^*\cdot[\delta \varepsilon]\textbf{F}_{n\parallel}-\textbf{D}_{n\perp}^*\cdot[\delta \varepsilon]^{-1}\textbf{C}_{n\perp}) \textbf{u}\cdot\textbf{dS},
\end{equation}

\noindent
where $[\delta \varepsilon]$ represents the step change in $[\epsilon]$ across the boundary, $\textbf{F}_n=\frac{1}{2}(\textbf{E}_{n+1}+\textbf{E}_{n-1})$, $\textbf{C}_n=\frac{1}{2}(\textbf{D}_{n+1}+\textbf{D}_{n-1})=[\epsilon]\textbf{F}_n$. Furthermore, the normal and perpendicular directions are measured with respect to the boundary, and the normalization (\ref{eq16}) still applies. In general, one needs to add up the effect of both these two terms (\ref{eq16}) and (\ref{eq16a}) in order to obtain the correct answer \cite{14c}, however, as discussed in the previous section, usually one effect may dominate \cite{23b}, under the above mentioned conditions.

\subsection{Conservation law}

Interestingly, a conservation law for total optical power may be found by summing up (\ref{eq14}) over all harmonics as $I(1)-I(0)=\sum_{n=-\infty}^{\infty}I_n(1)-I_n(0)$. After some algebra and dropping the terms which do no contrinute to the imaginary part, this will result in

\begin{equation}
\label{eq17}
I(1)-I(0)=\frac{\Omega}{2}\sum_{n=-\infty}^{\infty}\Im\iint_{\textrm{UC}}\textbf{E}_{n+1}^*\cdot[\delta \epsilon]\textbf{E}_n d^2r,
\end{equation}

\noindent
The right-hand-side of (\ref{eq17}) is proportional to frequency of elastic waves $\Omega$, and depending on its sign describes the total power withdrawn from or deposited into the acoustic wavefront during the process of nonlinear mixing with optical beam. This result also makes use of the supposed lossless property of the dielectrics, which comes as the real symmetry of the tensor $[\delta \epsilon]$. In practice, we have $\Omega$ being much smaller than $\omega$ by five to six orders of magnitude, hence, in other words to a very reasonable approximation, the total optical power is a conserved quantity while it is divided and continuously being exchanged among various harmonics according to

\begin{equation}
\label{eq18}
I(1)-I(0)\approx 0.
\end{equation}

\subsection{Resonance conditions}
\subsubsection{Optomechanical waveguides}

In the infinitely-many coupled-mode equations (\ref{eq14}), only a few in practice dominate which satisfy the condition of resonant power exchange. All other modes out of resonance will be return the fetched power from other modes soon after propagation a few lattice constants along the waveguide. The reason becomes obvious by expansion of Bloch eigenmodes \cite{5} as

\begin{equation}
\label{eq19}
\textbf{E}_n(\textbf{r};\kappa_n)=e^{-i\kappa_n x}\textbf{e}(\textbf{r};\kappa_n)=e^{-i\kappa_n x}\sum_{G}\textbf{e}_{\kappa_n}(y;G)e^{-i G x},
\end{equation}

\noindent
in which $\kappa_n$ is the Bloch wavevector, $G=G_m=2\pi m$ are reciprocal lattice vectors, and $\textbf{e}(\textbf{r};\kappa_n)=\textbf{e}(\textbf{r}+{\hat x};\kappa_n)$ is the periodic envelope function of the $n$th harmonic. Similarly, we may use

\begin{equation}
\label{eq20}
[\delta \epsilon(\textbf{r})]=\Re[e^{-i\textrm{k}x}[\theta](\textbf{r};\textrm{k})],
\end{equation}

\noindent
with $\textrm{k}$ and $[\theta](\textbf{r};\textrm{k})$ respectively being the Bloch wavevenumber and envelope of the permittivity tensor, as a result of moving elastic wave.

Insertion of (\ref{eq19}) and (\ref{eq20}) in (\ref{eq15}) gives

\begin{equation}
\label{eq21}
\alpha_n\approx -\frac{\omega_n}{4} \Im \iint_{UC}d^2r \sum_{G,G',G''}(S_1+S_2+S_3+S_4),
\end{equation}

\noindent
where the expressions for $S_1$, $S_2$, $S_3$, and $S_4$ are given by

\begin{eqnarray*}
S_1 & = & e^{i(\kappa_n-\kappa_{n-1}-\textrm{k}+G-G'-G'')x}\\
& & \times \textbf{e}_{\kappa_n}^*(y;G)\cdot[\theta]_{n\textrm{k}}(y;G')\textbf{e}_{\kappa_{n-1}}(y;G''),
\end{eqnarray*}
\noindent
\begin{eqnarray*}
S_2 & = & e^{i(\kappa_n-\kappa_{n+1}-\textrm{k}+G-G'-G'')x}\\
& & \times \textbf{e}_{\kappa_n}^*(y;G)\cdot[\theta]_{n\textrm{k}}(y;G')\textbf{e}_{\kappa_{n+1}}(y;G''),
\end{eqnarray*}
\noindent
\begin{eqnarray*}
S_3 & = & e^{i(\kappa_n-\kappa_{n-1}+\textrm{k}+G+G'-G'')x}\\
& & \times \textbf{e}_{\kappa_n}^*(y;G)\cdot[\theta]^*_{n\textrm{k}}(y;G')\textbf{e}_{\kappa_{n-1}}(y;G''),
\end{eqnarray*}
\noindent
\begin{eqnarray*}
S_4 & = & e^{i(\kappa_n-\kappa_{n+1}+\textrm{k}+G+G'-G'')x}\\
& & \times \textbf{e}_{\kappa_n}^*(y;G)\cdot[\theta]^*_{n\textrm{k}}(y;G')\textbf{e}_{\kappa_{n-1}}(y;G'').
\end{eqnarray*}

\noindent
The coupled-mode equations can be further simplified for resonant modes which satisfy the momentum equation $\kappa_n-\kappa_{n-1}\pm\textrm{k}=2\pi l$ where $l$ is an integer. In that case, only one of the above integrals does not vanish while integrating along many unit cells, and the other three will decay to zero. Practically, the surviving modes will be limited only to $n=0$ and one of the harmonics $n=\pm1$ since it would be normally impossible to satisfy all resonant criteria at once for all other harmonics.

Let us take $S_4$ as the resonant non-vanishing term, in which the initial wave at $n=0$ with frequency $\omega$ and the elastic wave with frequency $\Omega$ propagate in the same direction and the up-converted optical mode with $n=1$ and frequency $\omega_1=\omega+\Omega$ is counter-propagating. Hence, the counter-propagating up-converted optical beam must have opposing symmetry to the initial input beam. This can be satisfied only if the corresponding symmetries to the $n=0$ optical wave and elastic wave are different, in such a way that the integral in (\ref{eq21}) over the transverse direction $y$ does not vanish.

This condition may be easily satisfied in dielectric photonic crystal waveguides made out of two-dimensional triangular lattice, by simply removing one row of air holes along the $\Gamma\textrm{M}$ direction, where both even and odd branches co-exist within the same frequency range across the photonic bandgap.

For this purpose, the recently proposed snow-flake optomechanical crystals \cite{24} with some modifications could possibly provide a natural basis background lattice if simply one row of air defects are removed. The snow-flake optomechanical crystals exhibit the remarkable property of having both full mechanical and H-polarzied optical bandgaps. Of course, one may need to carefully engineer the resulting waveguide to make sure about the existence of crossing even-odd guided optical branches as well as at least one even or odd mechanical branch. Another interesting option is the very recent triangular design of patterned membranes \cite{21b} which exhibit an outstanding near-lossless operation and prolonged mechanical lifetimes.

The relation (\ref{eq20}) will then simplify as

\begin{eqnarray}
\label{eq22}
\alpha_n\approx -\frac{2\pi\omega_n}{4} & &\Im \int_{-\infty}^{\infty}dy\\
\nonumber
& &\sum_{G,G'}\textbf{e}_{\kappa_n}^*(y;G)\cdot[\theta]^*_{n\textrm{k}}(y;G')\textbf{e}_{\kappa_{n-1}}(y;G''),
\end{eqnarray}

\noindent
where we note that $G$, $G'$, and $G''$ are not all independent at once, satisfying $G+G'-G''=2\pi l$ for some $l$. This would clearly correspond to the Umklapp process \cite{25}, which folds back the scattered waves with wavevectors outside the first Brillouin zone  unto the equivalent waves inside the first Brillouin zone.

Hence, there are four different resonance conditions which need to be satisfied at one to obtain a strong interaction between elastic and optical waves:
\begin{enumerate}
    \item Conservation of photon-phonon energies $\hbar\omega_n=\hbar\omega+n\hbar\Omega$,
    \item Conservation of momentum $\kappa_n-\kappa_{n\pm1}\pm\textrm{k}=G$ where $G=2\pi l$ is a reciprocal lattice vector,
    \item Matching transverse symmetry, that is all of the three initial and scattered optical wave and elastic wave are even, or exactly two out of these are odd,
    \item Matching polarization, so that the perturbation permittivity $[\delta\epsilon]$ caused by the elastic field, would be sensed both by the initial and scattered optical waves.
\end{enumerate}

Consideration of these four requirements, reveals that it is not possible to have both the initial and scattered optical modes on the same waveguiding dispersion curve. Instead, they may have opposite transverse symmetries and very close frequencies differing only by $\Omega$. For this purpose, the photonic crystal waveguide must have the two even and odd branches present and crossing at once, as discussed in the above.

This could be now easily exploited to design a non-reciprocal photonic crystal waveguide discussed in the above, where an odd acoustic wave is propagating in background along a preferred direction. Now, any incoming optical wave with even symmetry would be scattered into an odd mode, and of course with a tiny amount of shift in frequency. However, the same scattering could not be reversed in direction. If the even optical wave is coming from the opposite direction, scattering unto an odd wave would be impossible. Hence, the scattering criteria for waves coming from both directions could not be met at once. This is exactly the non-reciprocal operation.

There is an extra advantage of this design over other designs. First of all, the Umklapp process could be exploited to design many essentially different scatterings, such as optical even to even through mechanical even, optical even to odd through mechanical odd, optical odd to odd through mechanical even, and as such, which go beyond the first Brillouin zone. Secondly, an optically guided Gaussian pulse with relatively wide sprectrum in frequency and wavenumber could be completely scattered in whole, to another Guassian pulse. Equivalently, the phase matching criteria are not so restrictive, as fabrication imperfections as well as some spread in the acoustic spectrum would allow some degree of fuzziness over the mathematically accurate expressions.

\subsubsection{Optomechanical cavities}

In the case of optomechanical cavities, where no propagation occurs for any type of the waves, the momentum condition in the above is irrelevant and may be dismissed. Here, the energy exchange between optical modes will cause appearance of a small imagniary part in the frequencies as $\tilde\omega_n=\omega_n+i\beta_n$, where $\beta_n$ represents the power loss or intake rate for the $n$th mode. The master equation (\ref{eq11}) may be modified now as

\begin{equation}
\label{eq23}
\nabla\cdot \bar {\textbf S}_n= \Im[\tilde \omega_n(U_{en}+U_{mn})].
\end{equation}

\noindent
and then integrated over the entire $x-y$ plane to obtain

\begin{equation}
\label{eq24}
\iint \Im[\tilde \omega_n(U_{en}+U_{mn})] d^2r=0.
\end{equation}

\noindent
Separation of real and imaginary parts now gives

\begin{equation}
\label{eq25}
\frac{\beta_n}{\omega_n}=-\frac{\iint \Im[U_{en}+U_{mn}] d^2r}{\iint \Re[U_{en}+U_{mn}] d^2r}.
\end{equation}

\noindent
Now, plugging in (\ref{eq13}) gives

\begin{equation}
\label{eq26}
\beta_n=-\omega_n\frac{\iint\Im[\textbf{E}_n^*\cdot[\delta \epsilon](\textbf{E}_{n+1}+\textbf{E}_{n-1})] d^2r}{2\iint \Re[U_{en}+U_{mn}] d^2r},
\end{equation}

\noindent
for the rate $\beta_n$. Hence, the typical time-constant at which the energy is exchanged with the $n$th mode of the cavity is $\tau_n=\beta_n^{-1}$. It has to be noticed that while evaluating (\ref{eq26}), that it is subject to the normalization condition for electromagnetic energy of individual modes, equating the denominator to unity for all modes. Hence, we get

\begin{equation}
\label{eq27}
\beta_n=-\frac{\omega_n}{2}\iint\Im[\textbf{E}_n^*\cdot[\delta \epsilon](\textbf{E}_{n+1}+\textbf{E}_{n-1})] d^2r.
\end{equation}

\noindent
Again, apart from the requirements for conservation of photon-phonon energies, symmetry of modes, and matching polarizations, not every photonic cavity allows co-existence of two mode finely separated by $\Omega$. Hence, for modulation purposes, the quality factor of the cavity $Q$ must not exceed the ratio $\frac{\omega}{\Omega}$, while switching applications would require high quality factor cavities with $Q>>\frac{\omega}{\Omega}$, thus setting a typical lower limit of $Q$ to $10^5$ and higher for the optical spectrum.

\section{Conclusions}
We presented a rigorous approach to solve the optomechanical coupling equations between a time-harmonic elastic field and an optical wave. An infinite set of equations were resulted for individual harmonics, which because of strict resonance conditions would be practically limited to very few surviving modes. Expressions for the coupling rates and lengths were obtained in case of optomechanical waveguides. A conservation law was obtained which satisfied the preservation of total optical power and withdrawn power from the elastic field, however, it was discussed that the contribution from the latter part may be negligible when the frequencies of these two fields differ by many orders of magnitude, hence preserving the total optical power. We also discussed the extension of this theory for using in analysis of optomechanical cavities. Application examples to non-reciprocal light transmission inside an optomechanical waveguide, and also conceptual design of optomechanical switches were discussed.



\begin{thebibliography}{99}

\bibitem{1} J. D. Joannopolous, S. G. Johnson, J. N. Winn, R. D. Meade, \textit{Photonic Crystals: Molding the Flow of Light}, 2nd ed., Princeton University Press: Princeton, 2009.
\bibitem{2} A. Yariv, \textit{Quantum Electronics}, 3rd ed., John Wiley \& Sons: New York, 1989.
\bibitem{3} R. W. Dixon, ``Photoelastic Properties of Selected Materials and Their Relevance for Applications to Acoustic Light Modulators and Scanners,'' \textit{J. Appl. Phys.}, vol. 38, no. 13, pp. 5149-5153, 1967.
\bibitem{4} T. S. Narasimhamurty, \textit{Photoelastic and Electro-Optic Properties of Crystals}, Plenum Press: New York, 1981.
\bibitem{5} S. Khorasani, ``On the Bloch Theorem and Orthogonality Relations,'' \textit{ J. Laser Opt. Photonics}, vol. 2, 118, 2015.
\bibitem{6} Y.-Z. Wang, F.-M. Li, W.-H. Huang, X. Jiang, Y.-S. Wang, K. Kishimoto, ``Wave band gaps in two-dimensional piezoelectric/piezomagnetic phononic crystals,'' \textit{Int. J. Solids Struc.}, vol. 45, pp. 4203–4210, 2008.
\bibitem{7} E. J. Reed, M. Soljacic, J. D. Joannopoulos, ``Color of Shock Waves in Photonic Crystals,'' \textit{Phys. Rev. Lett.}, vol. 90, 203904, 2003.
\bibitem{8} J.-H. Lee , C.-Y. Koh, J. P. Singer, S.-J. Jeon, M. Maldovan, O. Stein, E. L. Thomas, ``25th Anniversary Article: Ordered Polymer Structures for the Engineering of Photons and Phonons,'' \textit{Adv. Mater.}, vol. 26, pp. 532–569, 2014.
\bibitem{9} A. R. Rezk, S. Walia, R. Ramanathan, H. Nili, J. Zhen Ou, V. Bansal, J. R. Friend, M. Bhaskaran, L. Y. Yeo, S. Sriram, ``Acoustic-Excitonic Coupling for Dynamic Photoluminescence Manipulation of Quasi-2D $\textrm{MoS}_{2}$ Nanoflakes,'' \textit{Adv. Mater.}, vol. 3, pp. 888-894, 2015.
\bibitem{10} D. A. Golter, T. Oo, M. Amezcua, K. A. Stewart, H. Wang, ``Optomechanical Quantum Control of a Nitrogen-Vacancy Center in Diamond,'' \textit{Phys. Rev. Lett.}, vol. 116, 143602, 2016.
\bibitem{11} T. K. Paraiso, M. Kalaee, L. Zang, H. Pfeifer, F. Marquardt, O. Painter, ``Position-squared coupling in a tunable photonic crystal optomechanical cavity,'' \textit{Phys. Rev. X},  vol. 5, 041024, 2015.
\bibitem{12} J. T. Hill, A. H. Safavi-Naeini, J. Chan, O. Painter, ``Coherent optical wavelength conversion via cavity optomechanics,'' \textit{Nat. Commun.}, vol. 3, 1196, 2012.
\bibitem{13} M. Aspelmeyer, T. J. Kippenberg, F. Marquardt, eds., \textit{Cavity Optomechanics - Nano- and Micromechanical Resonators Interacting with Light}, Springer: Berlin, 2014.
\bibitem{14} M. Aspelmeyer, T. J. Kippenberg,  F. Marquard, ``Cavity optomechanics,''
 \textit{Rev. Mod. Phys.}, vol. 86, pp. 1391-1452, 2014.
\bibitem{14a} C. M. Reinke, A. A. Jafarpour, B. Momeni, M. Soltani, S. Khorasani, A. Adibi, ``Nonlinear finite-difference time-domain simulation of $\chi^{(2)}$ and $\chi^{(3)}$ effects in two-dimensional photonic crystals,'' \textit{J. Lightwave Tech.}, vol. 24, no. 1, pp. 624-634, 2006.
\bibitem{14b} A. Naqavi, M. Miri, K. Mehrany,  S. Khorasani, ``Extension of Unified Formulation for the FDTD Simulation of Nonlinear Dispersive Media,'' \textit{IEEE Photon. Tech. Lett.}, vol. 12, no. 16, pp. 1214-1216, 2010.
\bibitem{14c} B. Djafari-Rouhani, S. El-Jallal, Y. Pennec, ``Phoxonic crystals and cavity optomechanics,'' \textit{Comptes Rendus Physique}, vol. 17, no. 5, pp. 555-564, 2016.
\bibitem{14d} J. Gomis-Bresco, D. Navarro-Urrios, M. Oudich, S. El-Jallal, A. Griol, D. Puerto, E. Chavez, Y. Pennec, B. Djafari-Rouhani, F. Alzina, A. Martínez, C. M.
Sotomayor Torres, ``A one-dimensional optomechanical crystal with a complete phononic band gap,'' \textit{Nat. Commun.}, vol. 5, 4452, 2014.
\bibitem{15} A. R. Rezk, J. K. Tan, L. Y. Yeo, ``HYbriD Resonant Acoustics (HYDRA),'' \textit{Adv. Mater.}, vol. 28, pp. 1970-1975, 2016.
\bibitem{16} M. H. Aram, S. Khorasani, ``Novel Variational Approach for Analysis of Photonic Crystal Slabs,'' \textit{Mater. Res. Exp.}, vol. 2, no. 5, 056201, 2015.
\bibitem{17}  P. Liu, Y. W. Zhang, H. J. Gao, ``Interior and Edge Elastic Waves in Graphene,'' \textit{J. Appl. Mech.}, vol. 80, no. 4, 040901, 2013.
\bibitem{18} V. M. Pereira, R. M. Ribeiro, N. M. R. Peres, A. H. Castro Neto, ``Optical Properties of Strained Graphene,'' Europhysics letters 92, no. 6, 67001, 2010.
\bibitem{19}  M. Oliva-Leyva, G. G. Naumis, ``Anisotropic AC Conductivity of Strained Graphene,'', \textit{J. Phys.: Cond. Matter}, vol. 26, 125302, 2014.
\bibitem{20} G.-X. Ni, H.-Z. Yang, W. Ji, S.-J. Baeck, C.-T. Toh, J.-H. Ahn, V.-M. Pereira, B. Ozyilmaz,``Tuning Optical Conductivity of Large-Scale CVD Graphene by Strain Engineering,'' \textit{Adv. Mater.}, vol. 26, no. 7, 1081, 2013.
\bibitem{21}  R. A. Norte, J. P. Moura, S. Groblacher, ``Mechanical Resonators for Quantum Optomechanics Experiments at Room Temperature,'' \textit{Phys. Rev. Lett.}, vol. 116, 147202, 2016.
\bibitem{21a} E. Serra, M. Bawaj, A. Borrielli, G. Di Giuseppe, S. Forte, N. Kralj, N. Malossi, L. Marconi, F. Marin, F. Marino, B. Morana, R. Natali, G. Pandraud, A. Pontin, G. A. Prodi, M. Rossi, P. M. Sarro, D. Vitali, M. Bonaldi, ``Microfabrication of large-area circular high-stress silicon nitride membranes for optomechanical applications,'' \textit{AIP Advances}, vol. 6, 065004, 2016.
\bibitem{21b} Y. Tsaturyan, A. Barg, E. S. Polzik, A. Schliesser, ``Ultra-coherent nanomechanical resonators via soft clamping and dissipation dilution,'' \textit{arXiv preprint}, 1608.00937, 2016.
\bibitem{21c} R. Van Laer, B. Kuyken, D. Van Thourhout, R. Baets, ``Interaction between light and highly confined hypersound in a silicon photonic nanowire,'' \textit{Nat. Photon.}, vol. 9, pp. 199-203, 2015.
\bibitem{21d} E. N. Glytsis, T. K. Gaylord, ``Rigorous three-dimensional coupled-wave diffraction analysis of single and cascaded anisotropic gratings,'' \textit{J. Opt. Soc. Am. A}, vol. 4, no. 11, pp. 2061-2080, 1987.
\bibitem{21e} H. A. Haus and W. Huang, ``Coupled-mode theory,'' \textit{Proc. IEEE}, vol. 79, no. 10, pp. 1505–1518, 1991.
\bibitem{21f} W. Suh, Z. Wang, S. Fan, ``Temporal Coupled-Mode Theory and the Presence of Non-Orthogonal Modes in Lossless Multimode Cavities,'' \textit{IEEE J. Quant. Electron.}, vol. 40, no. 10, pp. 1511-1518, 2004.
\bibitem{22} M. Eichenfield, J. Chan, R. Camacho, K. J. Vahala, O. Painter, ``Optomechanical Crystals,'' \textit{Nature}, vol. 462, pp. 78-82, 2009.
\bibitem{23} S. G. Johnson, M. Ibanescu, M. A. Skorobogatiy, O. Weisberg, J. D. Joannopoulos, Y. Fink, ``Perturbation theory for Maxwell’s equations with shifting material boundaries,'' \textit{Phys. Rev. E}, vol. 65, 066611, 2002.
\bibitem{23b} K. C. Balram,	M. I. Davanço,	J.-D. Song,	K. Srinivasan, ``Coherent coupling between radiofrequency, optical and acoustic waves in piezo-optomechanical circuits,'' \textit{Nat. Photon.}, vol. 10, pp. 346-352, 2016.
\bibitem{24} A. H. Safavi-Naeini, J. T. Hill, S. Meenehan, J. Chan, S. Groeblacher, O. Painter, ``Two-dimensional phononic-photonic bandgap optomechanical crystal cavity,'' \textit{Phys. Rev. Lett.}, vol. 112, 153603, 2014.
\bibitem{25} K. Sakoda, \textit{Optical Properties of Photonic Crystals}, 2nd ed., Springer: Berlin, 2005.
\end{thebibliography}
\end{document}